\documentclass[11pt]{article}
\usepackage{apacite}
\usepackage{enumitem}
\usepackage{amsmath}
\usepackage{amsfonts}
\usepackage{amssymb}
\usepackage{upgreek}
\usepackage{textgreek}
\title{Cooperative Speech, Semantic Competence, and AI\footnote{Acknowledgements...}}
\author{Mahrad Almotahari\footnote{{\sf malmotah@ed.ac.uk}}
\\
University of Edinburgh
\\
School of Philosophy, Psychology and Language Sciences}

\begin{document}
\maketitle
\begin{abstract}
\noindent Cooperative speech is purposive. From the speaker’s perspective, one crucial purpose is the transmission of knowledge. Cooperative speakers care about getting things right for their conversational partners. This attitude is a kind of respect. Cooperative speech is an ideal form of communication because participants have respect for each other. And having respect within a cooperative enterprise is sufficient for a particular kind of moral standing: we ought to respect those who have respect for us. Respect demands reciprocity. I maintain that large language models aren’t owed the kind of respect that partly constitutes a cooperative conversation. This implies that they aren’t cooperative interlocutors, otherwise we would be obliged to reciprocate the attitude. Leveraging this conclusion, I argue that present-day language models are incapable of assertion and that this raises an overlooked doubt about their semantic competence. One upshot of this argument is that knowledge of meaning isn’t just a subject for the cognitive psychologist. It’s also a subject for the moral psychologist. 
\end{abstract}
\section{Q\&A}
\noindent Large language models (LLMs) are text-predicting systems whose outputs are so sensitive to the statistical distribution of tokens in actual human speech that they often give the impression of linguistic competence (Piantadosi 2024). GPT-4o and Claude 3.7 are prominent examples. They implement an approach to AI known as Deep Learning. This approach can be pursued in a variety of ways, but the common denominator is a multilayered, hence “deep”, neural network that’s “pre-trained” on a massive repository of text-based data, sometimes on the order of several petabytes. As a result, the network’s system of weights determines a multidimensional, context-sensitive representation of the patterns in actual human speech. Adjustments can then be made to improve performance on specific tasks. This process involves reinforcement learning: roughly, the supervised conditioning of certain outputs to maximize alignment with our values. 

Whether LLM speech really does manifest linguistic competence is a controversial question. But the controversy is too large to address in a single paper. Here I target one part of the much larger debate, beginning with whether LLMs can make assertions and ending with a challenge to their semantic competence. 

I use ‘assertion’ as an epistemic kind term. The type of speech it designates has a distinctive epistemic nature: it’s a mechanism for transmitting knowledge in a particularly direct way, as part of a collaborative inquiry. Some assertions fall short of achieving this function. They’re vulnerable to criticism. I’ll say more about this complication much later. For now, I’ll simply acknowledge that my use of ‘assertion’ rests partly on a terminological decision and partly on a substantive assumption. One might use the term to pick out a wider range of declarative speech. I would then use a different term to focus on the epistemic kind that interests me. Or one might doubt whether there is a speech act of the kind I'm targeting (Cappelen 2011). Addressing {\it this} doubt would require a separate paper, but the assumption I’m making (that there is a speech act of the relevant type) is hardly an outlier (Williamson 2000; Kelp 2018). And it would be surprising to learn that the semantic competence of LLMs hinged on its falsity.

One more framework-level assumption deserves mention: the nature of a kind is revealed in its paradigm instances. Just as chemists might restrict their focus to pure samples of gold, I’ll rely only on paradigmatically successful assertions. One might call an alloy ‘gold’ if its dominant component is the element Au, but there’s no direct route from an observation entirely about some impure gold to a substantive conclusion entirely about the nature of gold. Likewise, I assume there’s no direct route from an observation entirely about non-paradigmatic assertions to a substantive conclusion entirely about the nature of assertion (Ford 2008). I’ll have more to say about “paradigmatically successful assertions” as I proceed.

My initial question, then, is this: can LLMs transmit knowledge in the same way that we humans paradigmatically transmit knowledge, by asserting a proposition and thereby giving their word? \textendash No, paradigmatically human knowledge transmission requires cooperation, and LLMs are incapable of being cooperative.\footnote{How, then, do they transmit knowledge? LLMs might be like reliable clocks. If this were right, then they would be {\it indicators} rather than {\it informants}. “Roughly, the distinction is that between a person’s telling me something and my being able to tell something from observation of him” (Craig 1990, p. 35). But some forms of animal signaling don’t comfortably fit this distinction and they might provide a better model of LLM knowledge transmission. Mallory (2023) explores another possibility.} This might seem implausible. Doesn’t the department printer occasionally cooperate? \textendash Yes, but my answer involves a special sense of the term, one at work in the most influential account of communication that I’m aware of, namely, the account in Paul Grice’s classic paper, `Logic and Conversation’. Specific applications of this account in lexical semantics are controversial. But even critics of the Gricean tradition acknowledge ``the obvious role for collaboration in language use'' (Lepore and Stone 2015, p. 200). It’s this obvious role\textemdash nothing more\textemdash that my argument exploits.

Normal linguistic behavior is explicable in terms of the structure that joint rational activities share (Lewis 1969; Grice 1989; Stalnaker 1999). It’s widely assumed that this structure is governed by the Cooperative Principle: “Make your contribution such as is required, at the stage at which it occurs, by the accepted purpose or direction of the talk exchange in which you are engaged” (Grice 1989, p. 26).\footnote{The phrase ‘talk exchange’ is curious. Why does Grice use it when there are perfectly good, and perfectly ordinary, alternatives: ‘conversation’, ‘discussion’, and ‘speech’? His choice may not violate the maxim of Manner, but insofar as it elicits a quizzical response, it doesn’t seem to perfectly accord with it either. Cf., footnote 14.} To the extent that communicators are rational, their behavior fits the Principle (p. 28). And this isn't an accident. 

Accidental adherence to the Principle is fragile and therefore incapable of explaining stable patterns in normal human speech (as the Principle is intended to). It also falls short of being rational. Plausibly, one isn’t rational by luck. Going forward, then, I’ll assume that rational communicators have a disposition to conform to the Cooperative Principle. They generally tend to take the goals of their interlocutor into proper account and behave accordingly; otherwise, they wouldn’t reliably have an adequate conception of “the accepted purpose or direction of the talk exchange”. 

A core thesis of this paper is that LLM speech isn’t cooperative in the Gricean sense. I’ll argue for the thesis at length and then draw out some of its consequences (Sections 2-6). Near the end, I’ll rely on my pessimistic conclusion to raise an overlooked doubt about their semantic competence (Section 7). I raise this doubt partly to explain why the debate about AI assertion matters and partly to provoke attention from “LLM enthusiasts”. Nothing I say resolves the matter. My goal is just to forcefully raise the doubt and identify a constraint on future AI research.

The discussion ahead suggests that knowledge of meaning isn’t merely a subject for the cognitive psychologist; it’s also a topic for the moral psychologist. This sort of view isn’t unprecedented (Haugeland 1998, ch. 2; Ludlow 2013, ch. 4). But it’s certainly not prominent. I think it deserves a greater share of the spotlight. This paper explains why. 

\section{Quasi-Assertion}

\noindent There’s a small but growing literature on whether chatbots make assertions. Many authors say they do (Green 2010; Nickel 2013; Wheeler 2017, 2020; Freiman and Miller 2020; Kasirzadeh and Gabriel 2020; Kneer 2021; Glease 2022; Freiman 2024). As LLM behavior becomes more humanlike over time, this sort of enthusiasm is likely to intensify and spread. But it’s also likely to strike some as a complete nonstarter, ``because our very concept of assertion is tied to the notion of deciding to say something which does or does not mirror what you believe'' (Williams 1973, p. 146). And it’s not at all obvious that LLMs are, or will ever be, capable of making decisions and forming beliefs.

The same basic conceptual worry can be raised in other ways (Mallory 2023). For example, assertion is intentional speech; it seems to be a conceptual truth that intentional action is the upshot of a prior intention. If LLMs don’t have full-blooded intentions, then their output doesn’t really qualify as assertion.

These conceptual arguments underestimate how resourceful the opposing view is and consequently misidentify the central issue. An LLM enthusiast might grant that the concept {\it assertion} involves the concepts {\it decision}, {\it belief}, and {\it intention}. They might even grant that language models won’t be capable of deciding, believing, or intending. But they’ll maintain that, when all goes well, the output of a declarative sentence by an LLM transmits information in fundamentally the same way that you or I would by assertively uttering the sentence. Call the LLM’s speech ``quasi-assertion” if you like; the conceptual arguments don’t, by themselves, entail that quasi-assertion is an inhuman form of communication. 

The fact that an assertion is preceded by certain attitudes may well be epiphenomenal with respect to how recipients of the message retrieve knowledge from it. The mechanics of knowledge transmission might care only about whether a sentential output is reliably caused by an antecedent information state. Whether that state is a belief, properly so-called, or whether the corresponding process culminates in a decision/intention, properly so-called, might be theoretically irrelevant. 

Here’s one context in which this response is discernible: “If a verbal announcement on an airport loudspeaker constitutes an assertion when it is made by a human employee, why doesn’t the same verbal announcement on the same loudspeaker constitute an assertion when made by a computer? The function of the message, {\it the explanation of why subjects get knowledge from it}, and the phenomenology are the same in both cases” (Freiman and Miller 2020, p. 428, emphasis mine; cf., Green 2010; Nickel 2013; Wheeler 2020). From this perspective, the conceptual arguments offer nothing more than a terminological victory. If there were some independent reason to think that quasi-assertion is fundamentally different from assertion as a mode of knowledge transmission, then these conceptual arguments would be unnecessary. In the next section, I’ll sketch one such reason.

\section{Assertion}

\noindent In paradigm cases where a speaker asserts something and their interlocutor comes to know it, cooperation and “discursive respect” play a crucial role. I want to clarify this role and explain why nothing comparable is true of LLM speech. The upshot will be that LLMs are incapable of assertion.

Here, in outline, is my argument: 
\begin{enumerate}[noitemsep]
\item{Being a cooperative speaker requires having discursive respect.}
\item{Present-day LLMs are incapable of discursive respect.\footnote{This is my spin on John Haugeland’s memorable observation: “The trouble with artificial intelligence is that computers don’t give a damn” (1998, p. 47).}}
\item{So, present-day LLMs are incapable of cooperative speech.}
\item{Making an assertion requires the ability to transmit knowledge in a certain way, viz., the way that’s exemplified by paradigmatically successful assertions.}
\item{The ability to transmit knowledge in that way requires the ability to be a cooperative speaker.\footnote{My argument doesn’t take a side in the debate between reductionists and anti-reductionists about testimony. When I discuss claims (4) and (5) in Section 6, I’ll expand on this point.}}
\item{Therefore, present-day LLMs are unable to make assertions.}
\end{enumerate}
I don’t expect this outline to be fully comprehensible, much less convincing, from the start. But it should be clear that if we get all the way to (6) via (1)-(5), then reliance on quasi-assertion won’t help the LLM enthusiast. If quasi-assertion is similar enough to genuine assertion to depend on cooperation for successful knowledge transmission, then a straightforward corollary of (3)-(5), substituting ‘quasi-assertion’ for ‘assertion’, is that LLMs are unable to make quasi-assertions. If, however, quasi-assertion doesn’t depend on cooperation, then it’s fundamentally a different way of transmitting knowledge, not the way we humans paradigmatically do so. 

Perhaps this is a false dilemma. Suppose an LLM state stands to discursive respect in the way that quasi-assertion stands to assertion. Call this state discursive “quasi-respect”. By supposition, the state has the same theoretical significance as its counterpart but no inessential anthropomorphic associations. Just as we paradigmatically transmit knowledge in a spirit of discursive respect, LLMs do so by making quasi-assertions in a spirit of discursive quasi-respect. Have we made any real progress then?

The case for (2) will demonstrate that nothing is both similar enough to discursive respect to have the same theoretical significance and yet different enough to be realizable in a present-day LLM. For it’s essential to the theoretical significance of discursive respect that it confers on its realizer a normative status that current LLMs lack. Discursive quasi-respect is, therefore, a failed posit.

The discussion ahead expands on (1)-(6), articulating a neglected perspective in the philosophy of AI: humanlike communication owes about as much to attitudes that have their home in moral psychology as it does to attitudes that belong to cognitive psychology. From this perspective, it’s no easier to construct an artificial word-giver than it is to build an artificial locus of normativity. LLM enthusiasts might be able to accommodate this perspective. For reasons that will be made explicit, I doubt it. But before we can be confident, the perspective will need a clearer and more compelling formulation. In the next two sections, I’ll address three questions to which my argument immediately gives rise. What does ‘discursive respect’ mean? Why think it’s a requirement on being a cooperative speaker? And why do LLMs lack it? I’ll proceed in that order.

\section{Discursive Respect}

\noindent Discursive respect keys the speaker’s behavior to the conversational goal of their interlocutor, thereby facilitating joint success. It’s an instance of a more general attitude. I’ll describe this attitude and define discursive respect as a special case. More accurately, I’ll summarize Stephen Darwall’s influential account of the general attitude and then adapt it to fit my purpose. Reliance on Darwall’s account isn’t random. Grice and Darwall address different problems, but they share a Kantian outlook that makes this a natural starting point. 

There’s a way of regarding a thing that “consists, most generally, in a disposition to weigh appropriately in one’s deliberations some feature of the thing in question and to act accordingly” (Darwall 1977, p. 38). Notice that we have here one disposition with two aspects. The attitude in question is a disposition to (a) appropriately weigh something when deliberating about what to do and to (b) act in a way that’s recommended by that deliberation. Following Darwall, I’ll call this way of regarding a thing respect.\footnote{Darwall (2006, p. 131) criticizes the characterization of respect in Darwall (1977) because it fails to accommodate the thesis that respect is a “second-personal reason”, that is, a reason for someone to do something only insofar as they participate in a social relation that makes it possible for the object of respect to address them second-personally (e.g., “Hey you! Get off my foot!”). Referring to the passage I’ve quoted above, Darwall (2006) says, “This makes respect something one can realize outside of a second-personal relation; one need only adequately register a fact about or feature of someone: that she is a person” (p. 131). If readers are sympathetic to this criticism, then I invite them to amend the account as Darwall (2006) suggests.}

To say that persons as such are owed respect is to say that one ought to take the fact that so-and-so is a person into proper account when deliberating about actions that might affect them, and to proceed in the sort of way that such deliberation would dictate. There’s room for disagreement about what it means to properly take such a fact into account. But that’s a subject for ethicists, not one that needs to be settled for my purpose. 

In addition to respect for persons as such, there’s respect for persons and things as occupants of certain roles or as individuals who present themselves in certain ways. The respect one has for one’s friends, family, and fellows is distinguished by the ways in which the corresponding relations bear on what would count as weighing them appropriately in one’s deliberation. Occasionally, the attitude has a moral dimension, but not always: “A boxer talks of having respect for his opponent’s left hook and an adventurer of respecting the rapids of the Colorado. Neither regards the range of morally permissible actions as restricted by the things in question. Rather each refers to something which he fails to consider appropriately at his peril” (p. 40).

Respect should be distinguished from one of its close relatives: the admiration that virtue elicits (p. 39). Call this attitude esteem.\footnote{Darwall’s terminology is a little different. The attitude I call respect he calls “recognition respect”. The attitude I call esteem he often calls “appraisal respect” (2006, pp. 122-123). Neither Darwall nor I intend to capture ordinary usage completely. There are contexts in which it would be perfectly reasonable to use ‘respect’ (‘esteem’) to refer to esteem (respect).}  According to Darwall, “one may have [esteem] for someone without having any particular conception of just what behavior {\it from oneself} would be required or made appropriate by that person’s having the features meriting one’s [esteem]” (1977, p. 39). 

I might admire you as a stamp collector. Your knowledge of, and dedication to, the enterprise might be truly impressive. But given that I don’t care about stamps, and don’t have any special interest in their collectors as such, my esteem for you as a philatelist doesn’t really play a role in my decision-making. In contrast, as a disposition to deliberate appropriately about you and act accordingly, my respect for you constrains decision-making in determinate ways. 

The notion of respect involves “the idea that the object of respect (or some feature or fact regarding it) is such that the class of ‘eligible’ actions are restricted” (p. 40). This sort of restriction needn’t conflict with any background goal. In fact, there’s a term of praise for anyone whose background goals cohere with such restrictions: they have integrity. So, some respect-based restrictions on the will might not be burdensome.
 
Whereas the respect that facilitates success within a joint activity demands reciprocity, esteem must be earned.\footnote{Not all forms of respect demand reciprocity (e.g., respect for the rapids of the Colorado).} Imagine a joint endeavor the success of which depends on its participants—you and me—cooperating. Suppose you have respect for me as a participant in the activity. You’re disposed to (a) take certain facts about me into proper account when deliberating about your next move in our shared venture and to (b) act accordingly. I benefit from this.\footnote{Suppose that appropriate deliberation and action requires punishing me for a transgression. So, if you have respect for me, wouldn’t I be made worse off? In this case, the benefit to me is somewhat different. Having deliberated appropriately, you can justify the punishment. The justification puts me in a position to {\it accept} it: to see it as legitimate, as an opportunity for growth. To that extent, I do benefit.} Your respect for me, in this context, promotes our success by ensuring that, for your part, our project is cooperative. This benefit to me entails a restriction on the range of actions eligible for you. So, if I fail to reciprocate the respect you have for me as a participant in the activity, then I’m not contributing my fair share to the success of our undertaking. I’m guilty of “free riding”. In my preferred idiom, I’m “mooching” off you. And, presumptively, one ought not mooch.\footnote{Mooching doesn’t have to interfere with someone’s goals to be unfair. Mechanisms of social control are designed to harmonize unfairness with the attitudes of a population. They don’t thereby eliminate the unfairness; they mask it.} This is why respect within a joint undertaking, the success of which depends on cooperation, demands reciprocity: if we’re engaged in a project of the relevant sort, and one of us has respect for the other, then the other ought to have respect for the one; otherwise, they would be mooching.

One has discursive respect for one’s conversational partner if one is disposed to (a) properly take their conversational goal into account when deliberating about how to advance the exchange and to (b) act accordingly. In such cases, your interlocutor’s conversational goal partly regulates your verbal behavior. Often, the goal is to acquire some information in response to a question. If there aren’t any overriding considerations, one must make one’s contribution to the exchange relevant, informative, and true. More concisely, one is disposed to {\it get things right} for the recipient of the message. Recall, now, that Gricean cooperation also involves a disposition to get things right for your interlocutor. This means something specific within the framework. It's a disposition to conform to the conversational maxims: Quantity ({\it Be informative!}), Quality ({\it Speak knowledgeably!}), Relation ({\it Be relevant!}), and Manner ({\it Be perspicuous!}). As a result, the conditions for discursive respect are invariably satisfied. In fact, Gricean cooperation appears to {\it be} discursive respect (cf., Pettit 2021). Here, I simply rely on the weaker claim in premise (1).\footnote{One might {\it show} respect without {\it having} it. Can LLMs be insincere? \textendash Maybe.}

Deception, manipulation, and other forms of uncooperative speech are all too common, but the Gricean tradition treats the kind of rational exchange that takes place in ideal circumstances as fundamental. The hope is that we might understand non-ideal cases in terms of a breakdown in the principles that explain the ideal ones, not in terms of a completely distinct set of principles. A powerful thought lies behind this theoretical orientation: uncooperative speech is parasitic; it’s possible only against a background of regular cooperation (Lewis 1975; Grice 1989; Pettit 2021). And this is just one of several Kantian themes in the philosophy of Grice (cf., the discussion of false promises in the {\it Groundwork}).

A closely related Kantian theme is that cooperation involves a mutually beneficial arrangement, i.e., reciprocity: “Cooperation involves the idea of fair terms. Fair terms of cooperation specify an idea of reciprocity: all who are engaged in cooperation and who do their part as the rules and procedures require, are to benefit in an appropriate way…” (Rawls 1993, p. 16). My brief derivation of the principle that success-facilitating respect demands reciprocity was based on this Rawlsian point. The key claim was that, within a joint activity whose success depends on cooperation, unreciprocated respect means that one isn’t contributing one’s fair share to the success of the enterprise. 

At this stage, one might accuse me of equivocating between two senses of ‘cooperation’: an inflated sense relevant to social-contract theory in moral and political philosophy, and a deflated sense relevant to the philosophy of language. Let’s pause to consider the matter. 

The paradigm of a reciprocal relation is a contractual agreement. Contracts are meant to ensure that each party’s interests are satisfied. If you hold your end of the bargain, then I ought to hold mine as well. Happily, Grice expressed sympathy for a contractualist understanding of his framework: “I was attracted by the idea that observance of the Cooperative Principle and the maxims, in a talk exchange, could be thought of as a quasi-contractual matter, with parallels outside the realm of discourse” (1989, p. 29). And this wasn’t some quirky feature of his thought. There are three important similarities between conversations and contractual agreements: (i) both are undertaken to advance a more or less complicated structure of harmonizing aims (Clark 1996); (ii) their execution relies on preexisting or locally agreed-upon conventions of language (see Brennan and Clark 1996 on “conceptual pacts”); and (iii) they generate various rights and responsibilities, the contours of which depend on how interlocutors “negotiate” the rules of engagement as part of their engagement (Pecking and Garrod 2004, p. 172).\footnote{“...[I]t is only in certain contexts, say, when you and I are trying to work out what to believe together, that we have any standing to demand that we each reason logically, and even here that authority apparently derives from a moral or quasi-moral aspect: our having undertaken a common aim” (Darwall 2006, p. 14).}  

Suppose I spot a stranger at the train station who looks lost. Perhaps there’s a relatively weak obligation to address their puzzlement, since I know the station well and I’m not in a hurry. But, if I stop and ask the stranger whether they need help, the obligation suddenly seems a lot stronger. 
\begin{enumerate}[noitemsep]
\item[]{Me: Do you need help finding something?}
\item[]{Stranger: Yes, I do. I was wondering where I might\textemdash}
\item[]{Me: Oh, you misunderstood me. I was just curious. Best of luck!}
\end{enumerate}
\noindent My response here isn’t just disrespectful; it reveals incompetence as a communicator.\footnote{Not all forms of uncooperative speech reveal conversational infelicity.} According to Grice, “There is some sort of understanding (which may be explicit but which is often tacit) that, other things being equal, the transaction should continue in appropriate style unless both parties are agreeable that it should terminate. You do not just shove off or start doing something else” (1989, p. 29). If the exchange doesn’t end on terms agreeable to both parties, then there’s a kind of breach resembling the violation of a contract. And the error is a form of conversational infelicity.\footnote{The American sociologist Erving Goffman observed that one function of “small talk” at the end of a conversation is to allow interlocutors a way of tacitly agreeing to an inoffensive parting of ways (1982, p. 120).}
 
Gricean cooperation and discursive respect facilitate communicative success. That’s partly what their theoretical significance consists in. And it’s because of this that when one participant in the conversation satisfies the condition, the other should as well, otherwise they wouldn’t be doing their part in the joint enterprise. The theoretical significance of Gricean cooperation and discursive respect is, therefore, the source of the obligation to reciprocate. If a quasi-attitude has the same significance, then it too demands reciprocity.\footnote{The “contractualism” implicit in the picture I’ve been sketching explains why Grice often uses ‘talk exchange’ and ‘transaction’—terms that call to mind a marketplace where there’s a justified presumption of fair play and reciprocity (Darwall 2006, pp. 47-48).}

Even if the analogy between conversations and contracts stretches things in some cases, there’s no denying that cooperative conversation is a minimum condition for achieving a rational ideal. Cooperative communicators engage in a joint inquiry—a “conversation conducted to intellectual effect” (Pettit and Smith 1996, p. 376)—either as one of the primary investigators or as a collaborator who facilitates the investigation of another. It’s in this sort of context, where success hinges on working together, that the respect one has for another demands reciprocity: if interlocutors are engaged in a project of this sort, and one has discursive respect for the other, then the other ought to have discursive respect for the one.

Communication with very young children, pets, and the mentally disabled might involve discursive respect without the demand for reciprocal treatment from the child, pet, or the disabled interlocutor. But these exceptions prove the rule: communicative success hinges more on your ability to move things forward. You occupy the role of a caretaker (or something similar). Given the cognitive difference between you and your interlocutor, there’s nothing unfair about this distribution of labor and therefore no demand for reciprocity. These forms of communication aren’t Gricean conversations. One doesn’t expect the speech of a very young child, a pet, or a mentally disabled interlocutor to generate normal conversational implicatures—an indication that the Cooperative Principle isn’t in force. So, the way knowledge is transmitted by the child, etc., differs from the way knowledge is transmitted in paradigm cases, where the Cooperative Principle is in force. 

Let me quickly summarize the discussion. 
\begin{enumerate}[noitemsep]
\item{Being a cooperative speaker requires having discursive respect.}
\item{Present-day LLMs are incapable of discursive respect.}
\item{So, present-day LLMs are incapable of cooperative speech.}
\end{enumerate}
These claims raised several questions. What does ‘discursive respect’ mean? Why think it’s a requirement on being a cooperative speaker? And what reason is there to think that LLMs lack it? I’ve addressed the first two questions. In the next section, I’ll discuss the third and complete the case for (3).\footnote{(3) undermines a recent argument for the claim that LLM words refer to objects in our external environment (Mandelkern and Linzen 2024). Very briefly, the argument is that (i) LLMs are part of a speech community in which tokens of a referring term (‘Peano’) form a causal-historical chain back to a corresponding object (Peano) and (ii) LLMs have a “light-weight intention” (in my idiom, a “quasi-intention”) to use the term as part of the same causal-historical chain. But I take it to be a platitude that membership in a linguistic community enables cooperative speech with other members of that community. Given this platitude, (3) indicates that LLMs aren’t, in the relevant sense, part of any linguistic community. This will be relevant in footnote 22.}

\section{Reciprocity}

Psychological terms are permissive. In the right context, these would be perfectly natural descriptions:
\begin{enumerate}[noitemsep]
\item[]{The security system is {\it deliberating}; give it a second.}
\item[]{The thermostat is {\it cooperating} today; it {\it thinks} the room is less than 65 degrees.}
\item[]{The printer {\it says} the ink cartridge is low.}
\end{enumerate}
But it’s equally natural to think the italicized expressions denote properties distinct from the corresponding properties you and I exemplify. Psychological properties of the latter sort have a distinct theoretical significance.\footnote{Even an interpretationist such as Dennett acknowledges that we often “stretch” (his term) the meanings of psychological predicates when we talk about lower animals and simple artifacts (1996, p. 27).} They characteristically explain a subject’s behavior by rationalizing it from their point of view. The behavior is made intelligible partly by virtue of the subject’s rationality. One doesn’t need to assume that thermostats and so on are rational for these descriptions to render their behavior intelligible. And that’s a good indication that they offer a different kind of explanation: the properties they mention have a distinct theoretical significance. This difference plausibly makes for a difference in kind, because the nature of a psychological property is determined by the explanations in which it typically figures. The upshot is that ordinary uses of psychological vocabulary fail to carve at the joints. They can treat very different kinds of things (the internal states of thermostats on the one hand and the internal states of persons on the other) as if they were objectively the same. At some level of generality, they undoubtedly are. But, at that level, real divisions in nature are obscured.

The flexibility of psychological terms makes it much too easy for theorists to equivocate, and therefore more likely that their disputes will arise from merely terminological differences. It also makes it impossible to detect where the epistemic joints are just by focusing on what sounds most natural to say (and, as I indicated in Section 1, I’m using ‘assertion’ to carve a certain epistemic joint). Progress here requires circumspection. Rather than asking whether the internal processing and output of an LLM manifests a disposition to appropriately “deliberate about” and “act on” its interlocutor’s conversational goal, I’ll ask whether any LLM disposition has the theoretical significance that discursive respect has. This will enable us to circumvent the threat of verbal disputation. 

In the next section, I’ll unpack claims (4)-(6) and thereby describe the theoretical significance of discursive respect: the attitude partly explains the transmission of knowledge in paradigmatically successful cases of assertion. In the absence of discursive respect, knowledge isn’t transmitted via the giving of one’s word.\footnote{One might immediately object: “Even without discursive respect, a speech act can still transmit knowledge, provided that it’s sufficiently reliable. If LLMs become very accurate, their text outputs can transmit knowledge to the listener regardless of whether they have discursive respect for us.” But if it turns out that LLMs communicate knowledge in roughly the way macaques do, via reliable signaling, then the view I’m defending would be vindicated. Remember, the relevant question is whether LLMs transmit knowledge in a paradigmatically human way, via assertion. Earlier I suggested that LLM knowledge transmission is comparable to animal signaling.} In the previous section, I argued that when discursive respect plays this role within a joint activity it ought to be reciprocated. This means there are certain things one ought not do. So, we can answer the question I’ve asked by determining whether we’re obliged to behave in the ways we would be if LLMs transmitted knowledge by exercising a disposition (discursive “quasi-respect”) with the same theoretical significance. 

If an LLM has anything relevantly like discursive respect for us when we interact with it, then its conversational goal ought to constrain our deliberation and behavior in the way a conversational partner’s goal should. But are we so constrained? (I assume for the sake of argument that there’s a suitably deflationary sense of the term ‘conversational goal’ relative to which it’s true that LLMs have conversational goals. Recall the point about thermostats. If there’s no sense in which LLMs have conversational goals—not even “quasi-goals”—then we can hardly be obliged to weigh their goals appropriately when deliberating. For comparison: if there’s no god, then we can hardly be obliged to weigh its attitudes appropriately when deliberating. The assumption I’m making is, therefore, a concession to the other side.) The answer is no, it’s not the case that we ought to have anything relevantly like discursive respect for LLMs. It follows that they don’t have anything of the sort for us, otherwise we would be obliged to reciprocate. I’ll now defend this claim.

Suppose I end a conversation with you abruptly by just walking away while you happen to be mid-sentence. My behavior wouldn’t be compatible with discursive respect. Shoving off like that wouldn’t appropriately take account of your conversational goal. Naturally, I would owe you an explanation or an apology. But, if I were to end my back-and-forth with an LLM by closing the tab while it’s responding, I would owe it nothing. Accusing me of disrespect seems out of place, to put it mildly. These judgments suggest an implicit but widely shared understanding that LLMs don’t have anything sufficiently like discursive respect for us. If they did, we would owe it to them in return, which would entail that such behavior is at least presumptively objectionable. But it obviously isn’t. 

Earlier I argued that the source of the obligation to reciprocate discursive respect within a cooperative conversation is fairness. A failure to reciprocate would mean that, although your partner is holding up their end of the bargain and contributing to the joint undertaking’s success, you’re not holding up your end. LLMs simply don’t have the kind of dignity that would make certain ways of behaving unfair treatment of them.\footnote{Some respondents tell me they feel pressure to interact with chatbots politely. What are we to make of this feeling? Consider a parallel case: video-game enthusiasts with animal-friendly sensitivities tend to feel uncomfortable about aspects of gameplay that reward the simulation of animal abuse. These gamers aren’t confused about the nature of virtual animals; they don’t believe that a virtual dog is a real dog. Plausibly, their feeling of discomfort is rooted in a pretense whose behavioral and emotional effects aren’t restricted to the imagination—a pretense encouraged and sustained by the realism of their video game. In short, the pretense isn’t fully “quarantined” from the rest of their mental lives. Feeling pressure to treat chatbots politely is grounded, I suggest, in the same kind of “imaginative contagion” (Gendler 2006).} Shoving off from a conversation with one is, therefore, unobjectionable.

What must be true of something for it to have the kind of dignity that demands fair treatment and, more specifically, discursive respect? Many possibilities come to mind. Here, I’ll simply consider one. 

Something has wellbeing iff it can be made better or worse off. I’m inclined to think that there’s a specific type of vulnerability, or potential for being made worse off, associated with being the target of discursive disrespect. It can make one “feel small” or threaten one’s cherished self-understanding. Both conditions are psychically disruptive. If the disruption is sufficiently intense or persistent, it might constitute a harm. Creatures that can be harmed in this way are owed a specific kind of consideration. Plausibly, present-day LLMs aren’t owed discursive respect because it’s not possible for them to suffer this kind of harm. If they were made vulnerable in the specific way that calls for discursive respect, my reasoning in terms of reciprocity would vindicate the claim that they’re capable of Gricean cooperation. 

I’ve been arguing for premise (2) indirectly: being capable of discursive (quasi)respect means being capable of imposing an obligation on a conversational partner for reciprocal treatment; present-day LLMs are incapable of imposing an obligation of that sort because they’re not vulnerable in the way they would have to be to do so; therefore, etc. But LLM enthusiasts readily concede something from which (2) receives more straightforward support. For example, according to Mahowald et al. (2024), “LLMs routinely generate false statements, informally known as ‘hallucinations’. This observation is unsurprising: their training objective is to generate plausible sentence continuations, with no reference to the underlying factual correctness of the resulting claims” (p. 24). And this unsurprising observation has led some folks in the tech industry to conclude that LLMs are “bullshitters”, in the sense of Frankfurt (1998); they simply don’t care about the truth (y Arcas 2022). Bullshitting is, of course, incompatible with having discursive respect. But classifying LLMs as bullshitters may in fact be overly charitable. 

Paradigm bullshitters at the very least know how the truth or falsity of a sentence ought to regulate its use in assertoric speech. This is partly what makes bullshitting so reprehensible: despite understanding the significance of truth for communication, the bullshitter is indifferent to it. In Section 7, I’ll sketch an argument to the effect that if LLMs are incapable of discursive respect, then they don’t possess this understanding.\footnote{Browning (2023) offers a different take on LLM hallucination.} One obvious follow-up question, then, is: how can they be semantically competent in its absence? {\it That} is the challenge I advertised at the outset. Although we’re not quite ready for it, I can identify a nice feature of the reasoning ahead: it’s compatible with the view that present-day LLMs have all sorts of propositional attitudes, including “world knowledge” (Goldstein and Levinstein 2025). The line of thought I’m pursuing suggests that there’s something about {\it semantic} competence that's beyond their reach.

\section{Assertion Revisited}

\noindent I'll now consider the second part of my argument.
\begin{enumerate}[noitemsep]
\item[4.]{Making an assertion requires the ability to transmit knowledge in a certain way, viz., the way that’s exemplified by paradigmatically successful assertions.}
\item[5.]{The ability to transmit knowledge in that way requires the ability to be a cooperative speaker. }
\item[6.]{Therefore, present-day LLMs are unable to make assertions.}
\end{enumerate}
(4) and (5) jointly entail that making an assertion requires the ability to be a cooperative speaker. But one might think that very young kids and the mentally ill often make assertions, even though some are surely incapable of speaking cooperatively. Doesn’t this observation undermine the case for (6)?

No. There are two senses of ‘assertion’: the colloquial sense whose application isn’t restricted to an epistemic kind—it picks out any broadly declarative speech act—and the technical sense that is. There’s nothing wrong with describing some forms of speech as assertions (in the colloquial sense) even when they’re produced by someone who, for whatever reason, isn’t able to be a cooperative speaker. It’s no more problematic than grouping tomatoes with vegetables or jadeite with nephrite. Superficial similarities warrant the conflation for ordinary, non-theoretical purposes. But I’m not interested in the colloquial sense of ‘assertion’. I’m interested in the technical sense, for which I reserved the term at the very outset. The technical sense of ‘assertion’ does track an epistemic kind, specifically, the type of speech that’s supposed to transmit knowledge in a particularly direct way, as part of a joint inquiry in which, paradigmatically, Gricean cooperation facilitates success. Conversations with very young kids and the mentally ill aren’t Gricean in nature. We don’t {\it blame} the very young and the disabled for being uncooperative; we {\it manage} them (cf., Butlin and Viebahn 2024). This doesn’t mean they can’t transmit knowledge; it only means that they do so in a different way.

Together with (3), (4) and (5) justify the inference to (6). Crucially, these claims are neutral concerning the debate between reductionists and anti-reductionists in the epistemology of testimony. That debate is about what must be true of hearers for their testimony-derived beliefs to be justified. Reductionists deny that reliance on an epistemic principle unique to the case of testimony is necessary for the justification of a hearer’s testimony-derived belief. Equivalently, they maintain that testimony is not a basic source of justification, that a hearer’s testimony-derived beliefs aren’t presumptively justified. Anti-reductionists affirm the necessity of such reliance. They maintain that testimony is a basic source of justification, that a hearer’s testimony-derived beliefs are presumptively justified (Goldberg 2015, ch. 2). (4) and (5), however, are claims about the speaker’s side of the testimony relation. By themselves they leave the debate wide open.\footnote{Reductionism is compatible with the view that testimonial knowledge is epistemically distinctive; it just requires that the source of its distinctiveness be something other than the epistemic principles on which hearers rely to justify their testimony-derived beliefs (Goldberg 2006).} My argument neither requires nor rules out the view that a hearer’s testimony-derived beliefs are presumptively justified.  

Premise (4) shouldn’t be terribly controversial. To make an assertion is (at least in part) to invite your interlocutor to update their state of mind and thus the common ground in a certain way (Stalnaker 1999). If the content of the assertion is {\it p}, then you’re inviting them to believe {\it p}. If all goes well on your end, then the assertion was made knowledgeably and with the interlocutor’s conversational goal weighing as it should on your decision to speak. That is, the assertion was motivated by a concern for getting things right: addressing the question under discussion in a way that’s relevant, informative, and true. If all goes well on your interlocutor’s end, they’ll retrieve the content of your assertion, update accordingly, and acquire the relevant knowledge, thereby achieving their conversational goal. Examples with this structure are “paradigmatically successful assertions”. To make an assertion is to do your part in the joint effort to realize the structure, or to make it seem as if you’re doing your part. Liars and manipulators abound. In any event, you’re exercising an ability to transmit knowledge in the manner these “good cases” exemplify. Liars and manipulators deceive the unwitting partly because they exercise this ability, thereby falsely representing themselves as trustworthy: as doing their part to realize the structure of the paradigm cases.\footnote{Bald-faced lies are an exception. But that's why they pose a philosophical problem (Sorensen 2007). The source of the problem is the phenomenon.} They exploit the practice, the nature of which is revealed in examples where things go well (Williamson 2000; Ford 2008). When we focus on those examples, we observe that an assertion is the exercise of an ability to transmit knowledge in a certain way.

Even Freiman and Miller seem sympathetic to something like (4). Recall their argument: “If a verbal announcement on an airport loudspeaker constitutes an assertion when it is made by a human employee, why doesn’t the same verbal announcement on the same loudspeaker constitute an assertion when made by a computer? The function of the message, {\it the explanation of why subjects get knowledge from it}, and the phenomenology are the same in both cases” (2020, p. 428, emphasis added). In defense of AI assertion, they emphasize the way an audience obtains knowledge from a computer’s verbal output. This suggests that they, too, think of assertion as fundamentally a mode of knowledge transmission. 

This leaves premise (5). Focus again on paradigmatically successful assertions. They reveal an ability to transmit knowledge in a certain way. Well, in exactly what way is that? –The way in which joint activities generally succeed: by the participants exercising their ability to cooperate. Coming to learn of your interlocutor that they weren’t cooperative—that they didn’t have discursive respect for you—would undermine the justification their assertion might otherwise afford. Indifference to your goal just is a lack of respect, and respect is what greases the wheels of successful joint undertakings, including talk exchanges.  You might still be able to acquire knowledge from the speaker in this sort of case, for you might have independent evidence that their verbal behavior was reliable despite being uncooperative. This might even be a relatively common way in which knowledge is transmitted. I’m not making a descriptive claim about how things typically work; I’m making a normative claim about how they ought to work—or, equivalently, how they work in the paradigmatically successful cases. To think that cooperation and respect are epiphenomenal in such cases would be to think that there’s something exceptional about paradigm conversations: unlike other kinds of joint activity where things go well, cooperation and respect play no explanatory role in them. But then in what sense would these talk exchanges be paradigm conversations? What makes them the “good cases”, where things work as they ought to?

Claims (3)-(5) motivate a direct response to Freiman and Miller’s argument. They claim that if a person’s announcement constitutes an assertion, a computer’s announcement should as well, in large part because “the explanation of why subjects get knowledge from it” would be the same. But if the computer is a present-day language model, (3) entails that it’s not cooperative. So, it doesn’t transmit knowledge in the same way as the person’s announcement (assuming the person’s announcement isn’t a robotic performance of a mindless task). (4) and (5) then jointly entail that the computer isn’t making an assertion at all.

\section{Semantic Competence}

\noindent The way language is used is one indication (albeit fallible and indirect) of the way language is cognized. One component of language cognition is knowledge of meaning. By clarifying the relationship between assertion and semantic competence, and leveraging our pessimism about LLM assertion, we can bring the preceding discussion to bear on the question whether LLMs know English.

I’m going to argue that present-day LLMs don’t understand what it is for a declarative sentence of English to be true. That is, they don’t competently possess the concept {\it true-in-English}.\footnote{Their condition is analogous to Bert’s in one way and disanalogous in another (Burge 1979). Recall that Bert isn’t competent with the concept {\it arthritis} and therefore doesn’t understand what it is for someone to have the ailment. I say that LLMs are incompetent with {\it true-in-English} and therefore don’t understand what it is for an English declarative to have the value True. Still, Bert can think ‘arthritis’-thoughts by virtue of his membership in our linguistic community. But recall footnote 15. A corollary of (3) is that LLMs aren’t members of any linguistic community.} This conclusion  raises the doubt I advertised at the outset: how can an LLM know the meaning of an English declarative sentence if it doesn’t understand what it is for such a sentence to be true? I’m not going to argue that this challenge {\it can’t} be met; but reasonably maintaining that LLMs know English requires meeting it.

I'll begin by arguing for the following thesis:
\begin{enumerate}[noitemsep]
\item[7.]{Competence with the concept {\it true-in-English} suffices for knowing how to regulate assertoric speech in English to achieve the aim that speakers in paradigm conversations undertake when making an assertion.}
\end{enumerate}
Admittedly, (7) is far from obvious. But there’s an influential line of thought in the early work of Michael Dummett that both clarifies and motivates it. Various authors have expressed sympathy for Dummett’s line of thought,\footnote{See Field (1972), Glanzberg (2003), MacFarlane (2014), and Shaw (2014).} and it plays an important role in disagreements about the structure of natural-language semantics. Specifically, it imposes a constraint on the postulation of truth-value gaps and non-standard evaluation parameters. Even if you think there aren’t any such things, the hypothesis that there are ought to be intelligible. After all, whether it’s true is an empirical question (Glanzberg 2003; MacFarlane 2014; Shaw 2014). I take it to be a virtue of Dummett’s thought that it provides a way to understand these semantic posits and thereby facilitates {\it systematic} disagreement about them.\footnote{If there’s a kind of untruth that’s distinct from falsity, Dummett's line of thought indicates that it must correspond to a normative status for assertion distinct from both correctness and incorrectness, otherwise this allegedly distinct kind of untruth may just be a poorly labeled species of falsity. What could this third normative status for assertion be? Perhaps there isn’t one (Glanzberg 2003). Or perhaps there is (Shaw 2014). In any case, we now have a way to responsibly adjudicate the matter, by determining whether a third normative status is necessary for an adequate account of a certain observable practice.} 

I won’t claim that Dummett’s line of thought is irresistible. It obviously isn’t, since it relies on a substantive theory of conceptual competence. One might reasonably prefer a different theory of the subject. But it seems to me that any serious argument, which attempts to  vindicate or problematize LLM knowledge of meaning, will be theoretical in nature and therefore vulnerable in precisely the same way. Consider, for example, the influential case in Piantadosi and Hill (2022) for attributing semantic competence to modern language models on the basis of inferentialism about meaning. In this respect, there’s nothing uniquely problematic about my argument. 

Because (7) is a claim about conceptual competence, I’ll start by regimenting my talk of concepts. A concept is more than just a way of classifying things into distinct categories. In addition, it has a characteristic function in one’s mental life. Take a familiar example: the concept {\it winning-in-chess} isn’t merely a way of partitioning all possible end-game board configurations into two categories, those in which one color “wins” and those in which neither does (Dummett 1959, p. 143). It also serves a practical function, namely, regulating one’s gameplay. If a conceptually competent subject in a paradigm game judges that a particular move is likely to undermine her chances of winning, then (ignoring certain complications) she’s typically motivated to make a different move. The concept can perform this function because the subject knows that in a paradigm match the aim is to bring about a state of the board that belongs to the category of end-game configurations where her color “wins”. (There are unconventional ways of playing chess. A player might intend to lose. But conceptual competence hinges on knowledge of paradigm matches.) More generally, competence with a concept is knowing how to use the corresponding way of classifying things to perform its function and thereby achieve the characteristic aim one has in paradigmatic instances of the practice where the concept’s function is fixed.

The concept {\it true-in-English} is a way of classifying declarative sentences of the language into distinct categories (“the truths” and “the falsehoods”), a way of doing so that serves a particular function. According to Dummett (1959), the function is regulating the game of assertion. To competently possess the concept at issue is to know how to use the relevant way of classifying sentences to achieve the aim of asserting only truths. False assertions are criticizable and ought to be avoided. If a conceptually competent subject in a paradigm conversation judges that a sentence of English is false, then (ignoring certain complications) she’s typically motivated to avoid uttering it. 

Being blind to the way that {\it true-in-English} controls the practice of assertion is incompatible with understanding what it is for a sentence of English to be true. The analogy with {\it winning-in-chess}, and the general theory of conceptual competence the case of chess motivates, is the justification for (7). These considerations aren’t conclusive, but they confer on (7) a non-negligible degree of plausibility.

If LLMs are competent with {\it true-in-English}, then they know how to regulate assertoric speech in the required way. But regulating assertions in that way just is exercising a disposition to get things right for the recipient of the message. And to exercise {\it that} disposition is to manifest discursive respect. We saw in Section 5 that present-day LLMs are unable to do that. So, they’re unable to regulate assertoric speech in the required way. Can we somehow leverage this observation to make trouble for their semantic competence? I think we can, but it’s a little tricky. 

The inability to $\phi$ can coincide with knowledge how to $\phi$. For example, a violinist might know how to play the Kreutzer Sonata but be unable to do so because of a hand injury. Or someone might know how to speak English but be unable to put that knowledge to use on account of a brain lesion. These cases complicate the path forward. We can’t straightforwardly infer that present-day LLMs lack the know-how required for competence with {\it true-in-English} just because they instantiate a corresponding inability. But I think the complication can be finessed.

Cases in which practical knowledge coincides with a corresponding inability are cases where the removal of certain impediments\textemdash either naturally as part of normal human regeneration and growth or artificially by means of practice, medical intervention, or foul play\textemdash would suffice for realizing the ability. But the considerations that led us to conclude that present-day LLMs are unable to regulate assertoric speech in the required way indicate that it’s not by the subtraction of something present in them that they would be able to have discursive respect; it’s rather by the addition of something absent. What is this “something”?

Whatever it is, it has to ground the obligation for reciprocal treatment from an interlocutor. It is, therefore, a kind of dignity-conferring vulnerability. So, the inability of present-day LLMs to manifest discursive respect does support the conclusion that they don’t know how to regulate assertoric speech in the way required for competence with {\it true-in-English}. Just as the head of an academic department doesn’t know how to regulate the achievement of departmental aims if he doesn’t have professional respect for colleagues, one doesn’t know how to regulate assertoric speech for epistemic aims if one doesn’t have discursive respect for conversational partners. Given (7), it follows that present-day LLMs aren’t competent with the concept {\it true-in-English}. They don’t understand what it is for a declarative sentence to be true. How, then, can they know the meaning of such a sentence?

{\it Crucially}, one doesn’t have to think the meaning of a declarative sentence is its truth-conditions or referential content for this doubt to get a grip. It ought to bother you even if you think that the meaning of a declarative sentence is its inferential role. If the meaning of a sentence {\it is} its inferential role, then understanding a sentence’s inferential role should suffice for understanding its referential content and truth-conditions (Brandom 2000). After all, inferentialists aren’t {\it eliminativists} about reference and truth! In fact, LLM enthusiasts who rely on inferentialism to defend the attribution of semantic competence acknowledge the point here: that inferential role determines referential/truth-conditonal content (Piantadosi and Hill 2022, p. 2). How, then, can their enthusiasm withstand this doubt? More pointedly, should it? 
\pagebreak
\bibliographystyle{apacite}
\bibliography{Reference.bib}
\begin{enumerate}[wide, labelwidth=!, labelindent=-5pt, noitemsep]
\item[1.]{y Arcas, Blaise Agüera. 2022. ‘Do Large Language Models Understand Us?’ {\it Dædulus} 151: 
    183-197.}
\item[2.]{Brandom, Robert. 2000. {\it Articulating Reasons: An Introduction to Inferentialism}. Harvard University Press.}
\item[3.]{Brennan, Susan and Herbert H. Clark. 1996. ‘Conceptual Pacts and Lexical Choice in Conversation’ {\it Journal of Experimental Psychology} 22: 1482-1493.}
\item[4.]{Browning, Jacob. 2023. ‘Personhood and AI: Why Large Language Models Don’t Understand Us’ {\it AI $\&$ Society} 28: 1-8.}
\item[5.]{Burge, Tyler. 1979. `Individualism and the Mental' {\it Midwest Studies in Philosophy} 4: 73-121.}
\item[6.]{Butlin, Patrick and Emanuel Viebahn. 2024. ‘AI Assertion’ {\it Ergo}: Forthcoming.}
\item[7.]{Cappelen, Herman. 2011. ‘Against Assertion’, in J. Brown and H. Cappelen, eds., {\it Assertion: New Philosophical Essays}. Oxford University Press.}
\item[8.]{Clark, Herbert H. 1996. {\it Using Language}. Cambridge University Press.}
\item[9.]{Craig, Edward. 1990. {\it Knowledge and the State of Nature: An Essay in Conceptual Synthesis}. Oxford University Press.}
\item[10.]{Darwall, Stephen. 1977. ‘Two Kinds of Respect’ Ethics 77: 36-49.}
\item[11.]{Darwall, Stephen. 2006. {\it The Second-Person Standpoint: Morality, Respect, and Accountability}. Harvard University Press.}
\item[12.]{Dennett, Daniel. 1996. {\it Kinds of Minds: Toward an Understanding of Consciousness}. Basic Books.}
\item[13.]{Dummett, Michael. 1959. ‘Truth’ {\it Proceedings of the Aristotelian Society} 59: 141-162.}
\item[14.]{Field, Hartry. 1972. ‘Tarski’s Theory of Truth’ {\it Journal of Philosophy} 69: 347-375.}
\item[15.]{Ford, Anton. 2008. ‘Action and Generality’, in A. Ford, J. Hornsby, and F. Stoutland, eds., {\it Essays on Anscombe’s} Intention. Harvard University Press.}
\item[16.]{Frankfurt, Harry. 1998. {\it The Importance of What We Care About}. Cambridge University Press.}
\item[17.]{Freiman, Ori and Boaz Miller. 2020. ‘Can Artificial Entities Assert’. In S. Goldberg, ed., {\it Oxford Handbook of Assertion}. Oxford University Press.}
\item[18.]{Freiman, Ori. 2024. ‘AI-Testimony, Conversational AI and Our Anthropocentric Theory of Testimony’ {\it Social Epistemology} 38: 476-490.}
\item[19.]{Gendler, Tamar Szabó. 2006. ‘Imaginative Contagion’ {\it Metaphilosophy} 37: 183-203.}
\item[20.]{Glanzberg, Michael. 2003. ‘Against Truth-Value Gaps’, in J.C. Beal, ed., {\it Liars and Heaps}. Oxford University Press.}
\item[21.]{Glease, Amelia et al. 2022. ‘Improving Alignment of Dialogue Agents via Targeted Human Judgements’ {\it arXiv}:2209.14375.}
\item[22.]{Goffman, Erving. 1982. {\it Interaction Ritual: Essays on Face-to-Face Behavior}. Pantheon.}
\item[23.]{Goldstein, Simon and B.A. Levinstein. 2025. ‘Does ChatGPT Have a Mind?’ Unpublished Manuscript.}
\item[24.]{Goldberg, Sanford C. 2006. ‘Reductionism and the Distinctiveness of Testimonial Knowledge’, in J. Lackey and E. Sosa, eds., {\it The Epistemology of Testimony}. Oxford University Press.}
\item[25.]{Goldberg, Sanford. 2015. {\it Assertion: On the Philosophical Significance of Assertoric Speech}. Oxford University Press.}
\item[26.]{Green, Christopher. 2010. ‘Epistemology of Testimony’, in {\it Internet Encyclopedia of Philosophy}: https://iep.utm.edu/ep-testi/.}
\item[27.]{Grice, Paul. 1989. {\it Studies in the Way of Words}. Harvard University Press.}
\item[28.]{Haugeland, John. 1998. {\it Having Thought: Essays in the Metaphysics of Mind}. Harvard University Press.}
\item[29.]{Kasirzadeh, Atoosa and Iason Gabriel. 2023. ‘In Conversation with Artificial Intelligence: Aligning Language Models with Human Values’ {\it Philosophy of Technology} 36: 1-27.}
\item[30.]{Kelp, Chris. 2018. ‘Assertion: A Function First Account’ {\it Noûs} 52: 411-442.}
\item[31.]{Kneer, Markus. 2021. ‘Can a Robot Lie? Exploring the Folk Concept of Lying as Applied to Artificial Agents’ {\it Cognitive Science} 45: 1-15.}
\item[32.]{Lepore, Ernie and Matthew Stone. 2015. {\it Imagination and Convention: Distinguishing Grammar and Inference in Language}. Oxford University Press.}
\item[33.]{Lewis, David. 1969. Convention. Blackwell.}
\item[34.]{Lewis, David. 1975. ‘Languages and Language’, in K. Gunderson, ed., {\it Minnesota Studies in the Philosophy of Science} 7: 3-35.}
\item[35.]{Ludlow, Peter. 2011. {\it The Philosophy of Generative Linguistics}. Oxford University Press.}
\item[36.]{MacFarlane, John. 2014. {\it Assessment Sensitivity: Relative Truth and its Applications}. Oxford University Press.}
\item[37.]{Mahowald, Kyle et al. 2024. ‘Dissociating Language and Thought in Large Language Models’ {\it arXiv}:2301.06627v3.}
\item[38.]{Mallory, Fintan. 2023. ‘Fictionalism about Chatbots’ {\it Ergo} 10: 1082-1100.}
\item[39.]{Mandelkern, Matthew and Tal Linzen. 2024. ‘Do Language Models’ Words Refer?’ {\it Computational Linguistics} 50: 1191-1200.}
\item[40.]{Nickel, Philip J. 2013. ‘Artificial Speech and Its Authors’ {\it Minds and Machines} 23: 489-502.}
\item[41.]{Piantadosi, Steven. 2024. ‘Modern Language Models Refute Chomsky’s Approach to Language’, in E. Gibson and M. Poliak, eds., {\it From Fieldwork to Linguistic Theory: A Tribute to Dan Everett}. Berlin: Language Science Press.}
\item[42.]{Piantadosi, Steven and Felix Hill. 2022. ‘Meaning without Reference in Large Language Models’ {\it arXiv}:2208.02957v2.}
\item[43.]{Pettit, Philip. 2021. ‘A Conversive Theory of Respect’, in R. Deen and O. Sensen, eds., {\it Respect: Philosophical Essays}. Oxford University Press.}
\item[44.]{Pettit, Philip and Michael Smith. 1996. ‘Freedom in Belief and Desire’ {\it Journal of Philosophy} 93: 429-449.}
\item[45.]{Rawls, John. 1993. {\it Political Liberalism}. Columbia University Press.}
\item[46.]{Shaw, James. 2014. ‘What is a Truth-Value Gap?’ {\it Linguistics and Philosophy} 37: 503-534.}
\item[47.]{Sorensen, Roy. 2007. ‘Bald-faced Lies! Lying without the Intent to Deceive’ {\it Pacific Philosophical Quarterly} 88: 251-264.}
\item[48.]{Stalnaker, Robert. 1999. {\it Context and Content: Essays on Intentionality in Speech and Thought}. Oxford University Press.}
\item[49.]{Wheeler, Billy. 2017. ‘Giving Robots a Voice: Testimony, Intentionality, and the Law’, in S.J. Thompson, ed., {\it Androids, Cyborgs, and Robots in Contemporary Culture and Society}. IGI Global.}
\item[50.]{Wheeler, Billy. 2020. ‘Reliabilism and the Testimony of Robots’ {\it Techné} 24: 332-356.}
\item[51.]{Williams, Bernard. 1973. {\it Problems of the Self}. Cambridge University Press.}
\item[52.]{Williamson, Timothy. 2000. {\it Knowledge and Its Limits}. Oxford University Press.}
\end{enumerate}
\end{document}